Convolutional neural networks model improvements using demographics and image processing

filters on chest x-rays

Mir Muhammad Abdullah, Mir Muhammad Abdur Rahman, Mir Mohammed Assadullah

Abstract

**Purpose:** The purpose of this study was to observe change in convolutional neural networks (CNN) models' accuracies (ratio of correct classifications to total predictions) on thoracic radiological images by creating different binary classification models based on age, gender, and image pre-processing filters on 14 pathologies.

**Methodology:** This is a quantitative research exploring variation in CNN model accuracies. Radiological thoracic images were divided by age and gender and pre-processed by various image processing filters.

**Findings:** We found partial support for enhancement to model accuracies by segregating modeling images by age and gender and applying image processing filters even though image processing filters are sometimes thought of as information filters.

**Research limitations:** This study may be biased because it is based on radiological images by another research that tagged the images using an automated process that was not checked by a human.

**Practical implications:** Researchers may want to focus on creating models segregated by demographics and pre-process the modeling images using image processing filters. Practitioners developing assistive technologies for thoracic diagnoses may benefit from incorporating demographics and employing multiple models simultaneously with varying statistical likelihood.

**Originality/value:** This study uses demographics in model creation and utilizes image processing filters to improve model performance.

Convolutional neural networks model improvements using demographics and image processing filters on chest x-rays

Wang, et. al. (2017) published a database of 108,948 chest x-rays of 32,717 patients called "ChestX-ray14" on Kaggle.com. They obtained the images from a National Institute of Health (NIH) database and performed Natural Language Processing (NLP) on the associated reports to automatically classify each image with zero or more of twelve thoracic pathology keywords (Atelectasis, Cardiomegaly, Consolidation, Edema, Effusion, Emphysema, Fibrosis, Hernia, Infiltration, Mass, Nodule, Pleural Thickening, Pneumonia and Pneumothorax). Some rudimentary data about each image, such as gender and age, was also included in the compilation of the findings in a text file. This classification was not verified by a human and thus it may be incomplete or incorrect in certain cases. Our work is based on this data and thus our models may have been built on, hopefully slightly, inaccurate classifications.

This work extends the work of Wang, et. al. (2017) by applying deep convolutional neural networks (CNN) to sub-sets of these images sliced on age & gender and by applying some common image processing filters. We wanted to explore the areas where performance of CNN models is better in the subsets and when the image processing filters were applied. This should help determine the direction for fine-tuning these models in promising sub-sets and filters. In a recent article, Yamashita, Nishio, Do, & Togashi (2018) provide an overview of CNN and its applications in radiology.

Several researchers have also worked on the dataset published by Wang, et. al. (2017) notably Rajpurkar, et al., (2017), Zhou, Li, & Wang (2018), Yates, Yates, & Harvey (2018), Tang, et al., (2018) among others. We have not come across anyone using demographics and image processing filters to improve model performance on this dataset.

**Hypotheses**

Our experiments are broad in nature and we explored the general direction where additional research could be more fruitful. We anticipated that by dividing the patients by gender and by age we may be able to get better modeling results. For each of the twelve findings (Atelectasis, Cardiomegaly, Consolidation, Edema, Effusion, Emphysema, Infiltration, Mass, Nodule, Pleural Thickening, Pneumonia and Pneumathorax) we divided the images by gender and by age of those 54 years or older and those 53 years or younger. For each of the findings we had the following nine subsets: All, Males, Females, 53 years or younger, 54 years or older, 53 years or younger males, 54 years or older males, 53 years or younger females, 53 years or older females. We performed t-tests to see if there was any difference in modeling performance by dividing the population as such.

We also performed image processing filters on the raw images to see if that would have any impact on the modeling performance. Applying image processing filters is generally considered to be information filters because they alter the original information in the information. Nevertheless, we applied the filters of blur, edge detection, emboss, equalize, and sharpen, expecting better finding detection due to some of these filters. After applying these filters, we performed CNN modeling on all 14 thoracic findings, all eight subsets based on age and gender plus the control subset.

Our hypotheses are:

H1: CNN models would improve in accuracy if the sample data is divided by gender and age.

H2: CNN models would improve in accuracy if image processing filters are applied to the images prior to building the models.

## Research Methods

**Data Preparation**

 To focus on a finding, we ignored images that had multiple findings. We also ignored multiple images of the same person that were taken as follow up, to not bias the models. And finally, we ignored all images of those less than eighteen years of age.

 We divided the images for each condition by age of 53 years or less and 54 years or more, and by gender. For each of the subsets, we randomly selected 70% for training, 15% for validation, and 15% for testing, each of these subsets intermixed with randomly selected images labeled as "No Finding". For Atelectasis we created a subset of 2,059 images, for Cardiomegaly 521 images, for Consolidation 602 images, for Edema 302 images, for Effusion 1,921 images, for Emphysema 426 images, for Fibrosis 360 images, for Hernia 54 images, and for Infiltration 4,494 images, plus an equal number of random images with "No Finding".

 We processed all these images through image processing filters. All of the images were first processed with equalize and sigmoidal contrast filters. The equalize filter performs histogram equalization on the image for each channel. Sigmoidal contrast filter increases the contrast without saturating highlights or shadows. We chose a typical value of 3 for the sigmoidal transfer function and the mid-point of the maximum slope change to be 50% for middle-grey. We created a set of images with just these two filters. The options for the filters are moderate and commonly used. In addition to these two filters, we also applied blur, edge, emboss, and sharpen filters after applying equalize and sigmoidal filters.

 The blur filter convolves the image using a Gaussian function. We used the radius of 10 and standard deviation of 5 for this filter. The edge filter de tects an edge within a radius. We used the edge filter with a mild radius of 3 pixels. For the emboss filter we chose a radius of 3 and standard deviation of 1 for this filter. The sharpen filter sharpens the image using a Gaussian operation. We chose a radius of 3 with the standard deviation of 1 for this filter too.

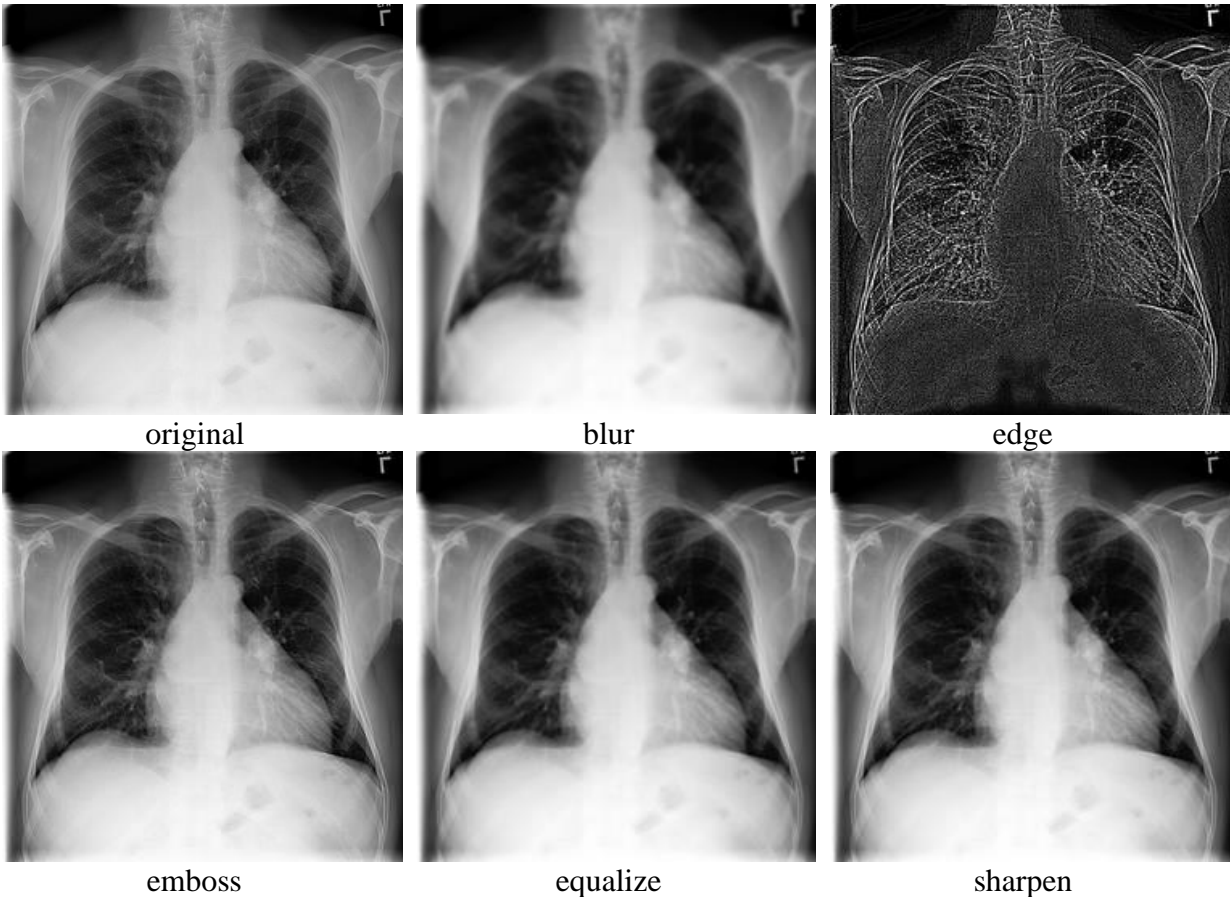

*Figure 1 Original image along with images after going through the five image processing filters of blur, edge, emboss, equalize, and sharpen.*

**Training Models**

We used NVIDIA's DIGITS (NVIDIA, 2018) version 6.0.0 running Caffe (Jia, et al.,

2014) version 0.15.14 in a Docker (Docker, 2018) container. We used black and white image

type of size 256 x 256 pixels that were squashed prior to modeling. We normalized the images

by subtracting a mean image from all images. For each configuration we created 10 models using

Stochastic Gradient Descent (SGD) solver over 30 epochs with a batch size of 50 images. The

learning rate was 0.01 with a Sigmoid Decay of 50% step and 0.1 Gamma. For CNN we used

AlexNet (Krizhevsky, Sutskever, & Hinton, 2012). It is a relatively simple network with five convolutional network layers and eight total network layers. From the results of the 30 epochs, we selected the highest accuracy model for that exercise to be used as input for downstream statistical analyses. Thus, each subset had ten selected models, each with a different random seed, each giving the highest accuracy among the 30 epochs of batches of 50 images. We calculated average and standard deviation of these highest accuracy models and used them to calculate p-values.

## Results

The general population was divided into 53 or less years of age, 54 or more years of age, males, females, males who are 53 years of age or less, males who are 54 years of age or more, females who are 53 years of age or less, and females who are 54 years of age or more. Table 1 presents the average accuracies for each of these demographic subsets and for each of the reported thoracic finding as the top number in the table cells. The bottom number is the difference of accuracy average from the general population for that reported finding. This average for the general population is reported in the column titled 'All' in the table. Positive changes with respect the 'All' column have green shade, whereas negative changes have red shade.

*Table 1* Average accuracies of subsets with change in average accuracy for each subset. In each cell, the top number refers to the average accuracy of the subset indicated in the column heading for the finding in the row label. The bottom number is the difference in average accuracy from the 'All' group of the entire set. * indicates the difference is significant at p-value <= 0.05 and ** indicates the difference is significant at p-value <= 0.01.

| Finding | All | 53- | 54+ | Males | Females | Males 53- | Females 53- | Males 54+ | Females 54+ |
|---|---|---|---|---|---|---|---|---|---|
| Hernia | 57.00 0.00 | 50.00 -7.00** | 54.40 -2.60 | 55.00 -2.00 | 65.00 8.00* | 95.00 38.00** | 50.00 -7.00** | 50.00 -7.00** | 63.60 6.60 |
| Nodule | 54.80 0.00 | 59.44 4.64** | 55.40 0.60 | 55.72 0.92* | 56.40 1.60** | 55.47 0.67 | 55.10 0.30 | 61.73 6.93** | 57.70 2.90** |
| Pneumo -thorax | 63.37 0.00 | 65.70 2.33** | 66.73 3.36** | 66.73 3.36** | 60.20 -3.17** | 67.50 4.13** | 63.93 0.56 | 56.60 -6.77** | 70.30 6.93** |
| **Atelectasis** | 66.89 0.00 | 69.37 2.48** | 63.80 -3.09** | 69.50 2.61** | 68.72 1.83* | 67.85 0.96 | 72.27 5.37** | 66.30 -0.59 | 67.73 0.84 |
| **Pleural Thicken.** | 61.20 0.00 | 67.90 6.70** | 55.00 -6.20** | 65.90 4.70** | 56.00 -5.20** | 63.60 2.40* | 70.20 9.00** | 68.80 7.60** | 52.00 -9.20** |
| **Infiltration** | 60.79 0.00 | 63.68 2.89** | 59.00 -1.79** | 60.38 -0.42* | 63.88 3.09** | 58.54 -2.25** | 64.53 3.73** | 64.50 3.71** | 56.56 -4.23** |
| **Emphysema** | 63.47 0.00 | 64.60 1.13 | 56.30 -7.17** | 62.40 -1.07 | 63.20 -0.27 | 67.80 4.33** | 59.40 -4.07* | 68.60 5.13** | 65.60 2.13 |
| **Mass** | 59.54 0.00 | 59.85 0.31 | 58.93 -0.61 | 64.25 4.71** | 55.93 -3.61** | 55.93 -3.61** | 61.40 1.86* | 60.90 1.36 | 59.20 -0.34 |
| **Effusion** | 75.70 0.00 | 79.80 4.10** | 75.30 -0.40 | 79.20 3.50** | 77.80 2.10** | 72.60 -3.10** | 74.27 -1.43** | 72.35 -3.35** | 69.87 -5.83** |
| **Pneumonia** | 62.00 0.00 | 59.40 -2.60* | 57.80 -4.20 | 58.20 -3.80** | 52.40 -9.60** | 58.40 -1.00 | 61.00 -1.00 | 65.40 3.40 | 71.80 9.80** |
| **Fibrosis** | 65.93 0.00 | 59.50 -6.43** | 63.50 -2.43** | 67.90 1.97* | 66.80 0.87 | 75.40 9.47** | 58.80 -7.13** | 59.40 -6.53** | 63.80 -2.13 |
| **Consolidation** | 82.05 0.00 | 70.33 -11.72** | 77.70 -4.35** | 73.27 -8.78** | 71.90 -10.15** | 78.50 -3.55** | 73.40 -8.65** | 75.40 -6.65** | 88.00 5.95** |
| **Edema** | 82.60 0.00 | 81.40 -1.20* | 85.80 3.20** | 76.20 -6.40** | 72.40 -10.20** | 77.00 -5.60** | 74.20 -8.40** | 76.20 -6.40** | 67.20 -15.40* |
| **Cardiomegaly** | 76.95 0.00 | 72.10 -4.85** | 68.50 -8.45** | 75.10 -1.85* | 65.10 -11.85** | 64.40 -12.55** | 65.00 -11.95** | 67.60 -9.35** | 63.00 -13.95** |

The change in average accuracy for Hernia among males of age 53 years or less is an outlier and it skewed the coloring scheme. The green color for this cell was eliminated to bring the rest of the positive results appropriately prominent. The table rows are sorted according the overall changes in average accuracies in descending order.

The table shows a mix of positive and negative results. Most positive changes by dividing the population as mentioned occurred in models for males of age less than 53 with hernia or fibrosis, females with hernia, females with age 53 years of less and pleural thickening and atelectasis, men and women of age 53 years or less and pleural thickening, males of age 54 years or more with nodule, pleural thickening, and emphysema, and females of age 54 years or older with hernia, pneumothorax, pneumonia, and consolidation. The subset models generally did not perform as well as the general population models for cardiomegaly, edema, consolidation (except for females of 54 years or older). Part of the reason is that cardiomegaly, edema, and fibrosis already had fairly good accuracy models for general population. That is the baseline for comparison for all the other populations segments.

Table 2 summarizes the results of preprocessing using image filters. The values are differences between the average accuracy of models with the filter applied for the condition in the column when subtracted from the average accuracy of models without the filter applied for the condition in the column. A single asterisk indicates a p-value less than or equal to 0.05 and a double asterisk indicates a p-value less than or equal to 0.01.

*Table 2* Average accuracies of the models. Inside the parenthesis is the difference in average accuracies of models created with image processing filters applied to images and without. * indicates p-value <= 0.05 and ** indicates p-value <= 0.01.

| Filter | Atelectasis | Cardiomegaly | Consolidation | Edema | Effusion |
|---|---|---|---|---|---|
| original | 66.89 (0.00) | 76.95 (0.00) | 82.05 (0.00) | 82.60 (0.00) | 75.70 (0.00) |
| blur | 67.14 (0.25) | 80.45 (3.50**) | 79.75 (-2.30**) | 80.80 (-1.80**) | 75.40 (-0.30) |
| edge | 63.95 (-2.94**) | 66.95 (-10.00**) | 77.75 (-4.30**) | 81.10 (-1.50*) | 76.03 (0.33) |
| emboss | 66.69 (-0.20) | 80.85 (3.90**) | 79.80 (-2.25**) | 81.00 (-1.60*) | 77.55 (1.85**) |
| equalize | 67.55 (0.66) | 81.55 (4.60**) | 79.45 (-2.60**) | 80.90 (-1.70*) | 76.13 (0.43) |
| sharpen | 67.49 (0.60) | 79.65 (2.70**) | 79.80 (-2.25**) | 81.30 (-1.30*) | 75.60 (-0.10) |

| Filter | Emphysema | Fibrosis | Hernia | Infiltration | Mass |
|---|---|---|---|---|---|
| original | 63.47 (0.00) | 65.93 (0.00) | 57.00 (0.00) | 60.79 (0.00) | 59.54 (0.00) |
| blur | 67.27 (3.80**) | 70.27 (4.33**) | 52.40 (-4.60**) | 58.11 (-2.68**) | 61.51 (1.97**) |
| edge | 61.47 (-2.00**) | 57.33 (-8.60**) | 53.40 (-3.60*) | 61.01 (0.22) | 55.20 (-4.34**) |
| emboss | 66.80 (3.33**) | 70.40 (4.47**) | 56.60 (-0.40) | 58.48 (-2.31**) | 60.34 (0.80) |
| equalize | 66.13 (2.67**) | 70.13 (4.20**) | 52.40 (-4.60**) | 58.51 (-2.28**) | 61.66 (2.11**) |
| sharpen | 65.13 (1.67*) | 67.27 (1.33*) | 53.60 (-3.40*) | 58.46 (-2.33**) | 60.77 (1.23) |

| Filter | Nodule | Pleural Thicken. | Pneumonia | Pneumothorax |
|---|---|---|---|---|
| original | 54.80 (0.00) | 61.20 (0.00) | 62.00 (0.00) | 63.37 (0.00) |
| blur | 55.50 (0.70) | 68.55 (7.35**) | 65.80 (3.80*) | 68.26 (4.89**) |
| edge | 57.58 (2.78**) | 57.85 (-3.35**) | 56.60 (-5.40**) | 70.11 (6.74**) |
| emboss | 55.95 (1.15**) | 67.95 (6.75**) | 66.60 (4.60**) | 69.83 (6.46**) |
| equalize | 55.98 (1.18*) | 67.80 (6.60**) | 64.20 (2.20) | 67.83 (4.46**) |
| sharpen | 55.38 (0.58) | 65.75 (4.55**) | 64.80 (2.80**) | 67.31 (3.94**) |

Green colored cells have shown an improvement in model performance, whereas orange and red colored cells have shown a degradation in the average accuracy of the models when the filters where applied. The bordered cells indicate the best filters for the thoracic finding written as the column header. Cardiomegaly, emphysema, fibrosis, pleural thickening, pneumonia, and pneumothorax findings generally improved applying the filters. Atelectasis, consolidation, edema, effusion, hernia, infiltration, mass, and nodule findings generally degraded by applying the image processing filters. The best filter for cardiomegaly was equalize increasing accuracy by 4.60%, for effusion it was emboss filter increasing average accuracy by 1.85%, for emphysema it was blur filter, for fibrosis it was emboss filter increasing average accuracy by 4.47%,  for mass it was equalize filter increasing by 2.11%, for nodule edge filter increased it by 2.78%, for pleural thickening blur filter increased it by 7.35%, for pneumonia emboss filter increased by 4.60%, and for pneumothorax edge filter increased the average model accuracy by 6.74%. No one filter seems to work for all thoracic findings. Pleural thickening, pneumothorax, fibrosis, and cardiomegaly findings responded positively to most filters.

Tables 3 through Table 7 report the findings of applying the five image processing filters of blur, edge, emboss, equalize, and sharpen to the population subsets according to age and gender mentioned earlier. Table 3 shows the results of applying blur filter to radiological images. In Table 3, the average accuracy increment for hernia among males of age 53 years or younger using the blur is again an outlier. The table cell showing the results has white background to not bias the remaining cells.

*Table 3* The blur filter was applied on the images prior to modeling. This table shows the average accuracies of subsets with change in average accuracy for each subset. In each cell, the top number refers to the average accuracy of the subset indicated in the column heading for the finding in the row label. The bottom number is the difference in average accuracy from the 'All' group of the entire set. * indicates the difference is significant at p-value <= 0.05 and ** indicates the difference is significant at p-value <= 0.01.

| Blur Finding | All | 53- | 54+ | Males | Females | Males 53- | Females 53- | Males 54+ | Females 54+ |
|---|---|---|---|---|---|---|---|---|---|
| Hernia | 52.40 0.00 | 54.40 2.00 | 52.00 -0.40 | 53.60 1.20 | 50.00 -2.40** | 80.00 27.60** | 60.00 7.60 | 52.60 0.20 | 56.20 3.80 |
| Atelec-tasis | 67.14 0.00 | 71.49 4.35** | 64.54 -2.60** | 70.03 2.89** | 71.68 4.54** | 67.65 0.51 | 77.00 9.86** | 64.90 -2.24** | 73.40 6.26** |
| Nodule | 55.50 0.00 | 54.92 -0.58 | 56.80 1.30* | 57.40 1.90** | 57.15 1.65** | 55.73 0.23 | 56.40 0.90 | 59.07 3.57** | 68.40 12.90** |
| Infiltra-tion | 58.11 0.00 | 61.94 3.83** | 59.04 0.93* | 57.93 -0.19 | 60.65 2.54** | 58.22 0.11 | 59.33 1.21** | 64.43 6.32** | 60.08 1.97** |
| Effusion | 75.40 0.00 | 77.23 1.83** | 74.40 -1.00** | 78.49 3.09** | 81.43 6.03** | 75.67 0.27 | 77.47 2.07** | 74.45 -0.95** | 71.40 -4.00** |
| Emphy-sema | 67.27 0.00 | 77.80 10.53** | 59.40 -7.87** | 63.30 -3.97** | 64.00 -3.27** | 69.20 1.93 | 64.80 -2.47 | 73.60 6.33** | 61.00 -6.27** |
| Mass | 61.51 0.00 | 61.45 -0.06 | 57.47 -4.05** | 66.95 5.44** | 54.47 -7.05** | 60.60 -0.91 | 54.00 -7.51** | 61.60 0.09 | 59.00 -2.51* |
| Pneumo thorax | 68.26 0.00 | 66.40 -1.86** | 70.87 2.61** | 65.53 -2.72** | 62.90 -5.36** | 73.40 5.14** | 63.47 -4.79** | 59.20 -9.06** | 62.50 -5.76** |
| Consoli dation | 79.75 0.00 | 69.60 -10.15** | 76.70 -3.05** | 70.47 -9.28** | 70.50 -9.25** | 78.60 -1.15* | 72.60 -7.15** | 80.20 0.45 | 84.00 4.25** |
| Edema | 80.80 0.00 | 79.80 -1.00 | 61.80 -19.00** | 75.80 -5.00** | 72.60 -8.20** | 89.20 8.40** | 79.60 -1.20 | 83.00 2.20 | 59.20 -21.60** |
| Pleural Thicken. | 68.55 0.00 | 64.60 -3.95** | 58.70 -9.85** | 69.40 0.85 | 59.50 -9.05** | 65.40 -3.15 | 64.60 -3.95** | 65.60 -2.95** | 55.20 -13.35** |
| Cardio-megaly | 80.45 0.00 | 77.30 -3.15** | 74.20 -6.25** | 75.70 -4.75** | 74.10 -6.35** | 80.40 -0.05 | 63.60 -16.85** | 74.40 -6.05** | 75.60 -4.85** |
| Pneu-monia | 65.80 0.00 | 65.20 -0.60 | 53.80 -12.00** | 64.20 -1.60 | 52.60 -13.20** | 57.80 -8.00** | 63.20 -2.60 | 55.20 -10.60** | 58.80 -7.00 |
| Fibrosis | 70.27 0.00 | 63.40 -6.87** | 61.30 -8.97** | 64.30 -5.97** | 70.40 0.13 | 63.00 -7.27** | 54.40 -15.87** | 56.20 -14.07** | 65.00 -5.27* |

This table also shows good improvements in average accuracy increment for models on female subjects, when broken down by age, for atelectasis by applying the blur filter. The best improvement was in the model detecting nodules among females of age 54 and up. The model for emphysema among 53 years old and younger also showed major improvement. Other notable improvements are for edema among males of 53 years of age or younger, infiltration among males, pneumothorax among males of age 53 years or younger, and consolidation among females of age 54 years or older.

Using blur filter to detect pleural thickening and fibrosis didn't fare well at all. The worse degradation in average accuracy occurred for some population segments when detecting edema, but the average accuracy for the baseline general population was already 80.80%. Likewise, cardiomegaly models also performed worse with blur filter, the baseline average accuracy was 80.45%. Consolidation models fared worse for almost all population segments because the baseline model for general population was already doing well at 79.75%.

Table 4 shows the results of applying the edge filter to the radiological images and then creating CNN models for various population segments. This filter application gave the best increment to the fibrosis models for all males, followed by pleural thickening model for males of age 53 years or less and females of age 53 years or less, and mass finding among females of age 54 years or older. Other notable increments are in detecting atelectasis among females of age 53 or younger, and mass detection for all of age 53 or younger. The baseline for edema was already at 81.10% and applying the edge filter made results only worse. Consolidation, pneumothorax, cardiomegaly, effusion, and emphysema detections also didn't improve much with the edge filter.

*Table 4* The edge filter was applied on the images prior to modeling. This table shows the average accuracies of subsets with change in average accuracy for each subset. In each cell, the top number refers to the average accuracy of the subset indicated in the column heading for the finding in the row label. The bottom number is the difference in average accuracy from the 'All' group of the entire set. * indicates the difference is significant at p-value <= 0.05 and ** indicates the difference is significant at p-value <= 0.01.

| Edge Finding | All | 53- | 54+ | Males | Females | Males 53- | Females 53- | Males 54+ | Females 54+ |
|---|---|---|---|---|---|---|---|---|---|
| Fibrosis | 57.33 0.00 | 63.30 5.97* | 63.10 5.77** | 70.00 12.67** | 63.80 6.47** | 59.60 2.27 | 52.00 -5.33** | 54.60 -2.73** | 55.80 -1.53 |
| Pleural Thicken. | 57.85 0.00 | 59.80 1.95** | 62.40 4.55** | 56.40 -1.45 | 53.50 -4.35** | 67.20 9.35** | 65.80 7.95** | 59.00 1.15 | 59.20 1.35 |

| Edge Finding | All | 53- | 54+ | Males | Females | Males 53- | Females 53- | Males 54+ | Females 54+ |
|---|---|---|---|---|---|---|---|---|---|
| Atelectasis | 63.95 0.00 | 66.14 2.19** | 62.29 -1.67** | 65.20 1.25* | 66.00 2.05** | 64.15 0.20 | 71.00 7.05** | 59.35 -4.60** | 66.73 2.78 |
| Hernia | 53.40 0.00 | 57.80 4.40 | 52.00 -1.40 | 50.00 -3.40** | 51.00 -2.40 | 60.00 6.60 | 55.00 1.60 | 56.60 3.20 | 50.00 -3.40** |
| Mass | 55.20 0.00 | 59.15 3.95** | 52.47 -2.73** | 55.50 0.30 | 52.40 -2.80** | 52.93 -2.27** | 55.20 0.00 | 55.50 0.30 | 63.20 8.00** |
| Infiltration | 61.01 0.00 | 60.09 -0.92** | 58.96 -2.05** | 59.71 -1.30** | 60.42 -0.60* | 59.76 -1.25** | 61.50 0.49 | 66.13 5.12** | 58.12 -2.89** |
| Nodule | 57.58 0.00 | 55.64 -1.94** | 53.40 -4.18** | 51.72 -5.86** | 53.00 -4.58** | 59.13 1.56 | 55.30 -2.28** | 60.00 2.43** | 60.80 3.22 |
| Pneumonia | 56.60 0.00 | 55.20 -1.40 | 54.00 -2.60* | 52.00 -4.60** | 52.40 -4.20** | 54.00 -2.60* | 53.80 -2.80 | 52.60 -4.00 | 60.00 3.40 |
| Emphysema | 61.47 0.00 | 61.20 -0.27 | 58.80 -2.67* | 59.20 -2.27** | 61.80 0.33 | 61.00 -0.47 | 52.00 -9.47** | 61.80 0.33 | 53.40 -8.07** |
| Effusion | 76.03 0.00 | 74.80 -1.23* | 71.30 -4.73** | 77.43 1.40** | 74.97 -1.07 | 69.93 -6.10** | 68.53 -7.50** | 73.05 -2.98** | 69.87 -6.17** |
| Cardiomegaly | 66.95 0.00 | 56.80 -10.15** | 67.70 0.75 | 68.10 1.15 | 64.60 -2.35 | 64.80 -2.15 | 63.40 -3.55 | 54.20 -12.75** | 59.80 -7.15** |
| Pneumothorax | 70.11 0.00 | 65.45 -4.66** | 65.67 -4.45** | 62.13 -7.98** | 64.90 -5.21** | 63.50 -6.61** | 62.27 -7.85** | 58.20 -11.91** | 69.50 -0.61 |
| Consolidation | 77.75 0.00 | 71.20 -6.55** | 76.60 -1.15 | 70.87 -6.88** | 59.80 -17.95** | 62.80 -14.95** | 62.00 -15.75** | 60.40 -17.35** | 70.00 -7.75* |
| Edema | 81.10 0.00 | 76.10 -5.00** | 74.20 -6.90** | 64.40 -16.70** | 70.80 -10.30** | 66.80 -14.30** | 71.60 -9.50* | 57.00 -24.10** | 57.80 -23.30** |

Table 5 demonstrates the improvements in average accuracy of models when the radiological images were first passed through the emboss filter. The clear outlier here, again, was the model detecting hernia among males of age 53 years or younger, and this cell was taken out of the color range. Applying the emboss filter increased the average accuracy of model detecting atelectasis for most population segments, especially females of age 53 years or younger and females of age 54 years or older. Interestingly, the average accuracy of all females when detecting atelectasis was 72.08% but when separate models were created for females of age 53 years and younger and females of age 54 years and older, the average accuracy went up to 76.60% and 75.33% respectively. Emphysema models also improved for all of age 53 years or younger, males of age 53 years or younger, and males of 54 years or older. Edema model for males 53 years or younger increased 6.40% to reach 87.40% average accuracy. Emboss filter didn't improve much accuracies for fibrosis, pneumonia, cardiomegaly, edema (except for the lone case mentioned of males of age 53 years or younger), pneumothorax, pleural thickening, and consolidation.

*Table 5* The emboss filter was applied on the images prior to modeling. This table shows the average accuracies of subsets with change in average accuracy for each subset. In each cell, the top number refers to the average accuracy of the subset indicated in the column heading for the finding in the row label. The bottom number is the difference in average accuracy from the 'All' group of the entire set. * indicates the difference is significant at p-value <= 0.05 and ** indicates the difference is significant at p-value <= 0.01.

| Emboss Finding | All | 53- | 54+ | Males | Females | Males 53- | Females 53- | Males 54+ | Females 54+ |
|---|---|---|---|---|---|---|---|---|---|
| Hernia | 56.60 0.00 | 52.60 -4.00 | 52.00 -4.60* | 55.00 -1.60 | 54.00 -2.60 | 85.00 28.40** | 70.00 13.40 | 60.00 3.40 | 56.40 -0.20 |
| Atelectasis | 66.69 0.00 | 72.83 6.14** | 64.89 -1.81** | 70.73 4.03** | 72.08 5.39** | 67.55 0.86 | 76.60 9.91** | 64.05 -2.64** | 75.33 8.64** |
| Infiltration | 58.48 0.00 | 61.85 3.37** | 58.55 0.06 | 59.76 1.28** | 60.35 1.87** | 56.86 -1.62** | 61.10 2.62** | 63.67 5.19** | 60.28 1.80** |
| Emphysema | 66.80 0.00 | 77.60 10.80** | 60.70 -6.10** | 65.30 -1.50 | 62.60 -4.20** | 73.20 6.40** | 70.60 3.80 | 72.10 5.30** | 60.60 -6.20** |
| Nodule | 55.95 0.00 | 54.88 -1.07* | 56.20 0.25 | 57.36 1.41** | 57.10 1.15* | 53.27 -2.68** | 53.30 -2.65** | 58.93 2.98** | 59.20 3.25** |
| Mass | 60.34 0.00 | 61.80 1.46** | 59.93 -0.41 | 64.05 3.71** | 55.20 -5.14** | 61.87 1.52* | 52.80 -7.54** | 64.10 3.76** | 59.30 -1.04 |
| Effusion | 77.55 0.00 | 79.20 1.65** | 75.70 -1.85** | 78.60 1.05** | 81.30 3.75** | 74.93 -2.62** | 77.07 -0.48 | 75.25 -2.30** | 70.47 -7.08** |
| Consolidation | 79.80 0.00 | 69.87 -9.93** | 77.70 -2.10** | 70.47 -9.33** | 71.00 -8.80** | 79.20 -0.60 | 72.00 -7.80** | 80.00 0.20 | 82.80 3.00** |
| Pleural Thicken. | 67.95 0.00 | 68.40 0.45 | 59.50 -8.45** | 70.50 2.55** | 61.40 -6.55** | 62.40 -5.55** | 59.60 -8.35** | 66.80 -1.15 | 56.20 -11.75** |
| Pneumothorax | 69.83 0.00 | 64.85 -4.98** | 69.60 -0.23 | 65.27 -4.56** | 63.80 -6.03** | 72.50 2.67** | 62.60 -7.23** | 54.20 -15.63** | 61.40 -8.43** |
| Edema | 81.00 0.00 | 81.60 0.60 | 63.60 -17.40** | 75.80 -5.20** | 69.00 -12.00** | 87.40 6.40** | 80.60 -0.40 | 78.60 -2.40 | 56.20 -24.80** |
| Cardiomegaly | 80.85 0.00 | 76.00 -4.85** | 72.30 -8.55** | 77.00 -3.85** | 72.60 -8.25** | 81.00 0.15 | 63.00 -17.85** | 71.60 -9.25** | 76.20 -4.65** |
| Pneumonia | 66.60 0.00 | 62.60 -4.00** | 56.40 -10.20** | 61.40 -5.20** | 54.40 -12.20** | 56.80 -9.80** | 60.80 -5.80 | 52.00 -14.60** | 67.20 0.60 |
| Fibrosis | 70.40 0.00 | 62.80 -7.60** | 61.50 -8.90** | 63.20 -7.20** | 70.40 -0.00 | 61.60 -8.80** | 54.60 -15.80** | 57.80 -12.60** | 62.00 -8.40** |

Before applying any filter, the equalize filter was applied to the images. Table 6 shows the results of applying just the equalize filter with no other filter applied to the images after it. In this table, we see the model accuracy increase for hernia on males with age 53 or younger being an outlier. It was taken out of the chloropleth coloring scheme to not bias colors of remaining cells. The greatest improvement came to emphysema model on those of age 53 years or less. The model of atelectasis for females of age 53 years or younger and 54 years or older, emphysema on males 53 years or younger and 54 years or older, infiltration on males 54 years or older, nodules on females 64 years or older also improved. Accuracy degradation occurred most for fibrosis and cardiomegaly, but their baseline accuracy was 70.13% and 81.55% respectively to begin with. Edema, pleural thickening, pneumonia, and consolidation all degraded in accuracy except consolidation among females of age 54 years or older and edema among males of age 53 years or younger.

*Table 6* The equalize filter was applied on the images prior to modeling. This table shows the average accuracies of subsets with change in average accuracy for each subset. In each cell, the top number refers to the average accuracy of the subset indicated in the column heading for the finding in the row label. The bottom number is the difference in average accuracy from the 'All' group of the entire set. * indicates the difference is significant at p-value <= 0.05 and ** indicates the difference is significant at p-value <= 0.01.

| Equalize Finding | All | 53- | 54+ | Males | Females | Males 53- | Females 53- | Males 54+ | Females 54+ |
|---|---|---|---|---|---|---|---|---|---|
| Hernia | 52.40 0.00 | 54.40 2.00 | 52.00 -0.40 | 53.40 1.00 | 54.00 1.60 | 80.00 27.60** | 60.00 7.60 | 56.40 4.00 | 55.60 3.20 |
| Atelectasis | 67.55 0.00 | 71.94 4.39** | 64.31 -3.24** | 71.05 3.50** | 72.44 4.89** | 67.80 0.25 | 75.80 8.25** | 65.60 -1.95** | 73.13 5.58** |
| Infiltration | 58.51 0.00 | 62.41 3.90** | 58.78 0.27 | 58.44 -0.07 | 60.30 1.79** | 58.06 -0.45* | 59.83 1.31** | 64.37 5.86** | 58.72 0.21 |
| Emphysema | 66.13 0.00 | 77.80 11.67** | 60.00 -6.13** | 65.50 -0.63 | 63.60 -2.53* | 72.00 5.87* | 64.40 -1.73 | 73.80 7.67** | 57.20 -8.93** |
| Nodule | 55.98 0.00 | 54.72 -1.26* | 55.45 -0.52 | 57.64 1.66** | 58.45 2.48** | 54.53 -1.44* | 53.40 -2.58** | 57.07 1.09* | 61.70 5.73** |
| Effusion | 76.13 0.00 | 77.90 1.77** | 76.20 0.07 | 79.94 3.81** | 81.13 5.00** | 74.80 -1.33** | 76.13 -0.00 | 75.00 -1.13** | 71.40 -4.73** |
| Mass | 61.66 0.00 | 61.55 -0.11 | 59.87 -1.79* | 66.35 4.69** | 54.80 -6.86** | 62.20 0.54 | 54.20 -7.46** | 63.70 2.04* | 58.80 -2.86** |

| Equalize Finding | All | 53- | 54+ | Males | Females | Males 53- | Females 53- | Males 54+ | Females 54+ |
|---|---|---|---|---|---|---|---|---|---|
| Pneumo thorax | 67.83 0.00 | 65.55 -2.28** | 69.67 1.84** | 64.87 -2.96** | 63.55 -4.28** | 71.90 4.07** | 62.67 -5.16** | 55.40 -12.43** | 65.50 -2.33** |
| Consoli dation | 79.45 0.00 | 70.27 -9.18** | 76.90 -2.55** | 70.60 -8.85** | 70.40 -9.05** | 78.90 -0.55 | 73.00 -6.45** | 80.00 0.55 | 82.80 3.35 |
| Pneumo nia | 64.20 0.00 | 63.00 -1.20 | 55.40 -8.80** | 63.20 -1.00 | 54.00 -10.20** | 57.80 -6.40** | 63.40 -0.80 | 61.20 -3.00 | 60.00 -4.20 |
| Pleural Thicken. | 67.80 0.00 | 68.20 0.40 | 59.30 -8.50** | 68.00 0.20 | 63.30 -4.50** | 61.40 -6.40** | 63.00 -4.80** | 65.00 -2.80** | 55.00 -12.80** |
| Edema | 80.90 0.00 | 79.80 -1.10* | 63.80 -17.10** | 75.80 -5.10** | 71.40 -9.50** | 87.60 6.70** | 79.20 -1.70** | 80.00 -0.90 | 61.80 -19.10** |
| Cardio megaly | 81.55 0.00 | 77.90 -3.65** | 76.00 -5.55** | 76.80 -4.75** | 73.50 -8.05** | 80.20 -1.35* | 60.40 -21.15** | 72.40 -9.15** | 79.40 -2.15* |
| Fibrosis | 70.13 0.00 | 61.70 -8.43** | 60.30 -9.83** | 68.20 -1.93* | 69.60 -0.53 | 63.40 -6.73** | 53.20 -16.93** | 58.20 -11.93** | 62.00 -8.13** |

Table 7 displays the results of applying sharpen filter to the radiological images. It had an amazing impact on the model for hernia among males 53 years or younger by increasing the average accuracy by 46.40% and making it 100%! Of course, this outlier was also taken out of the chloropleth color range to get a better view of other results in the table. Overall, this filter had positive impact on accuracies, especially atelectasis for females of age 53 years or less and females of 54 years or more, nodule among females of age 54 years or older and males of age 54 years or older, infiltration among 53 years or younger and males 54 years or older, emphysema among 53 years or younger and males 53 years or younger and males 54 years of age or older. Other notable improvements occurred in mass detection among males, edema among males of age 53 years or younger and 54 years or more, consolidation among females of age 54 years or older, and some improvement in pneumothorax among males of 53 years and younger.

*Table 7* The sharpen filter was applied on the images prior to modeling. This table shows the average accuracies of subsets with change in average accuracy for each subset. In each cell, the top number refers to the average accuracy of the subset indicated in the column heading for the finding in the row label. The bottom number is the difference in average accuracy from the 'All' group of the entire set. * indicates the difference is significant at p-value <= 0.05 and ** indicates the difference is significant at p-value <= 0.01.

| Sharpen Finding | All | 53- | 54+ | Males | Females | Males 53- | Females 53- | Males 54+ | Females 54+ |
|---|---|---|---|---|---|---|---|---|---|
| Hernia | 53.60 0.00 | 50.00 -3.60** | 52.00 -1.60 | 56.40 2.80 | 58.00 4.40 | 100.00 46.40** | 55.00 1.40 | 62.40 8.80 | 51.20 -2.40 |
| Atelectasis | 67.49 0.00 | 73.06 5.56** | 64.11 -3.38* | 70.43 2.93** | 70.88 3.39** | 67.60 0.11 | 76.93 9.44** | 64.90 -2.59** | 73.73 6.24** |
| Nodule | 55.38 0.00 | 54.88 -0.49 | 56.20 0.83 | 57.20 1.83** | 56.00 0.63 | 56.87 1.49* | 55.50 0.13 | 59.53 4.16** | 65.20 9.83** |
| Infiltration | 58.46 0.00 | 62.18 3.72** | 58.31 -0.15 | 58.58 0.12 | 60.50 2.04** | 58.26 -0.20 | 59.90 1.44** | 63.90 5.44** | 58.64 0.18 |
| Emphysema | 65.13 0.00 | 76.40 11.27** | 60.40 -4.73** | 60.70 -4.43** | 63.40 -1.73 | 72.40 7.27** | 66.60 1.47 | 72.30 7.17** | 61.00 -4.13 |
| Effusion | 75.60 0.00 | 77.73 2.13** | 76.73 1.13* | 78.94 3.34** | 79.77 4.17** | 74.80 -0.80* | 78.53 2.93** | 74.05 -1.55** | 70.13 -5.47** |
| Mass | 60.77 0.00 | 60.70 -0.07 | 58.40 -2.37** | 66.75 5.98** | 54.27 -6.50** | 61.53 0.76 | 53.30 -7.47** | 61.60 0.83 | 58.50 -2.27 |
| Pleural Thicken. | 65.75 0.00 | 65.20 -0.55 | 60.10 -5.65** | 69.30 3.55** | 64.10 -1.65* | 65.40 -0.35 | 63.20 -2.55** | 65.80 0.05 | 57.40 -8.35** |
| Pneumothorax | 67.31 0.00 | 64.90 -2.41** | 69.40 2.09** | 67.07 -0.25 | 63.30 -4.01** | 71.30 3.99** | 63.73 -3.58** | 58.00 -9.31** | 62.20 -5.11** |
| Consolidation | 79.80 0.00 | 69.80 -10.00** | 77.80 -2.00** | 69.47 -10.33** | 70.70 -9.10** | 78.30 -1.50* | 72.40 -7.40** | 80.00 0.20 | 85.60 5.80** |
| Fibrosis | 67.27 0.00 | 62.80 -4.47** | 63.00 -4.27** | 67.20 -0.07 | 69.40 2.13** | 62.80 -4.47** | 53.60 -13.67** | 58.20 -9.07** | 60.60 -6.67* |
| Cardiomegaly | 79.65 0.00 | 77.90 -1.75* | 75.30 -4.35** | 76.50 -3.15** | 72.00 -7.65** | 80.60 0.95 | 62.20 -17.45** | 74.20 -5.45** | 76.40 -3.25** |
| Edema | 81.30 0.00 | 80.50 -0.80 | 64.00 -17.30** | 75.60 -5.70** | 72.00 -9.30** | 88.00 6.70** | 80.60 -0.70 | 85.40 4.10** | 56.80 -24.50** |
| Pneumonia | 64.80 0.00 | 65.00 0.20 | 56.00 -8.80** | 63.60 -1.20 | 54.60 -10.20** | 58.40 -6.40** | 61.20 -3.60 | 54.40 -10.40** | 55.00 -9.80* |

The natural question now is, which filter works best for which combination of population segment and radiological finding. This is answered in Table 8. In each cell it has the highest average accuracy achieved, the difference from the original images for that subset and that finding (i.e. similar cell in Table 1). Here the color scheme is a bit different. Yellows are lower values and green are greater improvements in average accuracy for the finding in row label and population segment in column label. Among greatest increases in accuracy due to applying the filters are hernia models for females 53 years or younger using emboss filter increasing accuracy by 20% and males 54 years or older using sharpen filter increasing accuracy by 12.40%, cardiomegaly models for males 53 years or younger with emboss filter increasing accuracy by 16.60% and females 54 years or older with equalize filter increasing accuracy by 16.40%, edema models for males 53 years or younger with blur filter increasing accuracy by 12.20%, emphysema models for females 53 years or younger with emboss models increasing accuracy by 11.20%, nodule models for females 54 years or older using blur filter increase accuracy by 10.70%. Average model accuracy did not improve from the baseline original image models for 38 cases represented by the cells in Table 8.

*Table 8* This table shows the highest average accuracies of population subsets observed using the image processing filters. In each cell, the top number refers to the average accuracy of the subset indicated in the column heading for the finding in the row label. The number below it is the difference in average accuracy of the same population segment and finding of the highest average accuracy. * indicates the difference is significant at p-value <= 0.05 and ** indicates the difference is significant at p-value <= 0.01. The name of the filter that yielded the highest absolute accuracy is written below it. The cells are colored from yellow to green based on increase in average accuracy.

| Finding | All | 53- | 54+ | Males | Females | Males 53- | Females 53- | Males 54+ | Females 54+ |
|---------|-----|-----|-----|-------|---------|-----------|-------------|-----------|-------------|
| Atelecta sis | 67.55 (0.66) equalize | 73.06 (3.69**) sharpen | 64.89 (1.09**) emboss | 71.05 (1.55**) equalize | 72.44 (3.72**) equalize | 67.85 (0.00) original | 77.00 (4.73**) blur | 66.30 (0.00) original | 75.33 (7.60**) emboss |
| Cardiom egaly | 81.55 (4.60**) equalize | 77.90 (5.80**) equalize | 76.00 (7.50**) equalize | 77.00 (1.90*) emboss | 74.10 (9.00**) blur | 81.00 (16.60**) emboss | 65.00 (0.00) original | 74.40 (6.80**) blur | 79.40 (16.40**) equalize |
| Consoli dation | 82.05 (0.00) original | 71.20 (0.87) edge | 77.80 (0.10) sharpen | 73.27 (0.00) original | 71.90 (0.00) original | 79.20 (0.70) emboss | 73.40 (0.00) original | 80.20 (4.80**) blur | 88.00 (0.00) original |
| Edema | 82.60 (0.00) original | 81.60 (0.20) emboss | 85.80 (0.00) original | 76.20 (0.00) original | 72.60 (0.20) blur | 89.20 (12.20**) blur | 80.60 (6.40**) emboss | 85.40 (9.20**) sharpen | 67.20 (0.00) original |
| Effusion | 77.55 (1.85**) emboss | 79.80 (0.00) original | 76.73 (1.43**) sharpen | 79.94 (0.74) equalize | 81.43 (3.63**) blur | 75.67 (3.07**) blur | 78.53 (4.27**) sharpen | 75.25 (2.90**) emboss | 71.40 (1.53**) blur |

| Finding | All | 53- | 54+ | Males | Females | Males 53- | Females 53- | Males 54+ | Females 54+ |
|---|---|---|---|---|---|---|---|---|---|
| Emphysema | 67.27 (3.80**) blur | 77.80 (13.20**) blur | 60.70 (4.40**) emboss | 65.50 (3.10**) equalize | 64.00 (0.80) blur | 73.20 (5.40**) emboss | 70.60 (11.20**) emboss | 73.80 (5.20**) equalize | 65.60 (0.00) original |
| Fibrosis | 70.40 (4.47**) emboss | 63.40 (3.90**) blur | 63.50 (0.00) original | 70.00 (2.10) edge | 70.40 (3.60**) blur | 75.40 (0.00) original | 58.80 (0.00) original | 59.40 (0.00) original | 65.00 (1.20) blur |
| Hernia | 57.00 (0.00) original | 57.80 (7.80) edge | 54.40 (0.00) original | 56.40 (1.40) sharpen | 65.00 (0.00) original | 100.00 (5.00) sharpen | 70.00 (20.00*) emboss | 62.40 (12.40) sharpen | 63.60 (0.00) original |
| Infiltration | 61.01 (0.22) edge | 63.68 (0.00) original | 59.04 (0.04) blur | 60.38 (0.00) original | 63.88 (0.00) original | 59.76 (1.22**) edge | 64.53 (0.00) original | 66.13 (1.63**) edge | 60.28 (3.72**) emboss |
| Mass | 61.66 (2.11**) equalize | 61.80 (1.95**) emboss | 59.93 (1.00) emboss | 66.95 (2.70**) blur | 55.93 (0.00) original | 62.20 (6.27**) equalize | 61.40 (0.00) original | 64.10 (3.20**) emboss | 63.20 (4.00) edge |
| Nodule | 57.58 (2.78**) edge | 59.44 (0.00) original | 56.80 (1.40*) blur | 57.64 (1.92**) equalize | 58.45 (2.05**) equalize | 59.13 (3.67**) edge | 56.40 (1.30) blur | 61.73 (0.00) original | 68.40 (10.70**) blur |
| Pleural Thicken. | 68.55 (7.35**) blur | 68.40 (0.50) emboss | 62.40 (7.40**) edge | 70.50 (4.60**) emboss | 64.10 (8.10**) sharpen | 67.20 (3.60*) original | 70.20 (0.00) original | 68.80 (0.00) original | 59.20 (7.20**) edge |
| Pneumonia | 66.60 (4.60**) emboss | 65.20 (5.80**) blur | 57.80 (0.00) original | 64.20 (6.00**) blur | 54.60 (2.20) sharpen | 58.40 (0.00) original | 63.40 (2.40) equalize | 65.40 (0.00) original | 71.80 (0.00) original |
| Pneumothorax | 70.11 (6.74**) edge | 66.40 (0.70) blur | 70.87 (4.13**) blur | 67.07 (0.33) sharpen | 64.90 (4.70**) edge | 73.40 (5.90**) blur | 63.93 (0.00) original | 59.20 (2.60**) blur | 70.30 (0.00) original |

Table 9 may be of greater interest to practitioners as it contains the absolute best average accuracies, and color coded accordingly, for various models using different filters for a radiological finding. For instance, looking at the first row, if one is interested in detecting atelectasis, the average accuracy for general population is 67.55% after applying equalize filter. The average accuracy increases or decreases for different population segments, using the filters indicated in that cell.

Here we see a large swath of green cells towards the top of the table where model accuracies are mostly between 70% and 90%. The bottom three rows are light green and there the model accuracies are mostly between 60% to 70%. Systems to detect these conditions can show the likelihood of detecting a radiological finding based on the model performance using different filters for various population segments.

*Table 9* This table shows the highest average accuracies of population subsets observed using the image processing filters. In each cell, the top number refers to the average accuracy of the subset indicated in the column heading for the finding in the row label. The number below it is the difference in average accuracy of the same population segment and finding of the highest average accuracy. * indicates the difference is significant at p-value <= 0.05 and ** indicates the difference is significant at p-value <= 0.01. The name of the filter that yielded the highest absolute accuracy is written below it. The cells are colored from red to green based on absolute average accuracy.

| Finding | All | 53- | 54+ | Males | Females | Males 53- | Females 53- | Males 54+ | Females 54+ |
|---|---|---|---|---|---|---|---|---|---|
| Atelectasis | 67.55 (0.66) equalize | 73.06 (3.69**) sharpen | 64.89 (1.09**) emboss | 71.05 (1.55**) equalize | 72.44 (3.72**) equalize | 67.85 (0.00) original | 77.00 (4.73**) blur | 66.30 (0.00) original | 75.33 (7.60**) emboss |
| Cardiomegaly | 81.55 (4.60**) equalize | 77.90 (5.80**) equalize | 76.00 (7.50**) equalize | 77.00 (1.90*) emboss | 74.10 (9.00**) blur | 81.00 (16.60**) emboss | 65.00 (0.00) original | 74.40 (6.80**) blur | 79.40 (16.40**) equalize |
| Consolidation | 82.05 (0.00) original | 71.20 (0.87) edge | 77.80 (0.10) sharpen | 73.27 (0.00) original | 71.90 (0.00) original | 79.20 (0.70) emboss | 73.40 (0.00) original | 80.20 (4.80**) blur | 88.00 (0.00) original |
| Edema | 82.60 (0.00) original | 81.60 (0.20) emboss | 85.80 (0.00) original | 76.20 (0.00) original | 72.60 (0.20) blur | 89.20 (12.20**) blur | 80.60 (6.40**) emboss | 85.40 (9.20**) sharpen | 67.20 (0.00) original |
| Effusion | 77.55 (1.85**) emboss | 79.80 (0.00) original | 76.73 (1.43**) sharpen | 79.94 (0.74) blur | 81.43 (3.63**) blur | 75.67 (3.07**) blur | 78.53 (4.27**) sharpen | 75.25 (2.90**) emboss | 71.40 (1.53*) blur |
| Emphysema | 67.27 (3.80**) blur | 77.80 (13.20**) blur | 60.70 (4.40**) emboss | 65.50 (3.10**) equalize | 64.00 (0.80) blur | 73.20 (5.40**) emboss | 70.60 (11.20**) emboss | 73.80 (5.20**) equalize | 65.60 (0.00) original |

| Finding | All | 53- | 54+ | Males | Females | Males 53- | Females 53- | Males 54+ | Females 54+ |
|---|---|---|---|---|---|---|---|---|---|
| Fibrosis | 70.40 (4.47**) emboss | 63.40 (3.90**) blur | 63.50 (0.00) original | 70.00 (2.10) edge | 70.40 (3.60**) blur | 75.40 (0.00) original | 58.80 (0.00) original | 59.40 (0.00) original | 65.00 (1.20) blur |
| Hernia | 57.00 (0.00) original | 57.80 (7.80) edge | 54.40 (0.00) original | 56.40 (1.40) sharpen | 65.00 (0.00) original | 100.00 (5.00) sharpen | 70.00 (20.00*) emboss | 62.40 (12.40) sharpen | 63.60 (0.00) original |
| Infiltration | 61.01 (0.22) edge | 63.68 (0.00) original | 59.04 (0.04) blur | 60.38 (0.00) original | 63.88 (0.00) original | 59.76 (1.22**) edge | 64.53 (0.00) original | 66.13 (1.63**) edge | 60.28 (3.72**) emboss |
| Mass | 61.66 (2.11**) equalize | 61.80 (1.95**) emboss | 59.93 (1.00) emboss | 66.95 (2.70**) blur | 55.93 (0.00) original | 62.20 (6.27**) equalize | 61.40 (0.00) original | 64.10 (3.20**) emboss | 63.20 (4.00) edge |
| Nodule | 57.58 (2.78**) edge | 59.44 (0.00) original | 56.80 (1.40*) blur | 57.64 (1.92**) equalize | 58.45 (2.05**) equalize | 59.13 (3.67**) edge | 56.40 (1.30) blur | 61.73 (0.00) original | 68.40 (10.70**) blur |
| Pleural Thicken. | 68.55 (7.35**) blur | 68.40 (0.50) emboss | 62.40 (7.40**) edge | 70.50 (4.60**) emboss | 64.10 (8.10**) sharpen | 67.20 (3.60*) edge | 70.20 (0.00) original | 68.80 (0.00) original | 59.20 (7.20**) edge |
| Pneumonia | 66.60 (4.60**) emboss | 65.20 (5.80**) blur | 57.80 (0.00) original | 64.20 (6.00**) blur | 54.60 (2.20) sharpen | 58.40 (0.00) original | 63.40 (2.40) equalize | 65.40 (0.00) original | 71.80 (0.00) original |
| Pneumothorax | 70.11 (6.74**) edge | 66.40 (0.70) blur | 70.87 (4.13**) blur | 67.07 (0.33) sharpen | 64.90 (4.70**) edge | 73.40 (5.90**) blur | 63.93 (0.00) original | 59.20 (2.60**) blur | 70.30 (0.00) original |

## Conclusions

We started out wanting to find ways to boost accuracies of CNN models for thoracic findings and hypothesized the following two hypotheses.

H1: CNN models would improve in accuracy if the sample data is divided by gender and age.

H2: CNN models would improve in accuracy if image processing filters are applied to the images prior to building the models.

Table 1 shows some support for H1. Average accuracies for 62 (55.35%) sets of models for a thoracic finding on a particular population segment actually degraded from the baseline of all population segments mixed together from 112 such combinations of thoracic findings and populations segments. On the flip side, creating models for specific population segments increased accuracy for 50 out of 112 model sets, i.e. 44.64%. Although H1 was not completely or even overwhelmingly supported, there is some benefit towards increased accuracies in some population segments. In other words, age and gender segregations have some role in model performance and this can be exploited by practitioners.

Table 2 shows us the results of applying image processing filters on the general population while looking for the 14 thoracic pathologies. Applying image processing filters alters the original image and may hide some of the information present in it. At the same time, for certain cases, it may enhance the image make it easier for the downstream CNN to correctly classify the images. In our case, for 29 model sets out of 70 combinations of thoracic findings and image filters, i.e. 41.42% model sets, the average model accuracy decreased. Whereas, for 41 model sets, i.e. 58.57%, the accuracy increased. This supports the idea that for certain image classification problems it is possible that image processing filters may increase model accuracies despite the alteration to the original image.

Table 9 also gives us some insight into the support for H2. For 38 sets of models out of 126 total, i.e. 30.15% model sets there was no improvement in average accuracies from the baseline of similar cells in Table 1 (of original images). For the remaining 88 sets of models, i.e. 68.75% model sets there was improvement in average accuracy, 59 of them statistically significant at p-value of 0.01 and five of them at p-value of 0.05. In other words, 50.08% of all combinations of thoracic pathologies and population segments had statistically significant (p-value ≤ 0.05) increment in average accuracies. We find H2 also partially supported.

For 38 model sets out of 126 in Table 9, i.e. 30.15%, none of the five filters could improve average model accuracies above the baseline accuracies. This is in line with the belief that image processing filters filter out information. However, there was the unexpected finding in this study of roughly half the cases improving model accuracy due to some image processing filter. Apparently, for some cases the images become easier to classify after applying image processing filters. Amazingly, Gaussian blur filter was the best performing filter for 27 out of 126 model sets, i.e. 21.42%, in Table 9. This is very counterintuitive as the blur only blurs an image.

## Implications

For researchers, preprocessing images with various image processing filters, or incorporating such filters within the network, may be of value in certain applications. It opens up another area of research to increase overall model performance. This study employs a relatively simple network, AlexNet (Krizhevsky, Sutskever, & Hinton, 2012), with five convolutional layers to achieve the accuracies reported. However, the findings of this research can be applied to the works of several other researchers who have employed more sophisticated networks such as DenseNet (Huang, Liu, Weinberger, & van der Maaten, 2016) with 121 convolution layers or specialized networks for chest x-rays such as CheXNet (Rajpurkar, et al., 2017). Additionally, applying different filters for different demographic populations and different pathologies may also enhance the overall model performance.

For practitioners, the state of technology in Computer Aided Detection (CADe) and Computer Aided Diagnosis (CADx) of thoracic findings may not be appropriate for such an unsupervised system. At the same time, a complementary system that assists the radiologist such as suggesting top five findings may not be too far in future. These suggested top findings may come from different models based on different demography and image filters, each supported by statistical likelihood. Likewise, a system that confirms a radiologist's finding with a modeled finding to reduce radiology practice liability may also have value to radiology practices. Seeing varying accuracies based on age, gender, and image processing filters, it is important for those working on creating such systems to assist to utilize multiple models based on demography and expected findings. Such systems may then continually learn and improve their own models as radiologists share the findings with them.

**Limitations and future work**

The work by Wang et. al. (2017) is based on processing radiologists' findings using Natural Language Processing (NLP), where were never checked by a human. All of the models above are based on that work and may be impacted by any error in detecting those thoracic findings.

The present work is independent of the CNN network, as it focuses on increase the accuracy of models created by a network, in this case AlexNet (Krizhevsky, Sutskever, & Hinton, 2012). AlexNet is a relatively less complicated network, consisting of only five layers in the network. The same methodology can be applied to increase model accuracies using other networks such as GoogLeNet (Szegedy, et al., 2014) , a 22-layer network and DenseNet (Huang, Liu, Weinberger, & van der Maaten, 2016), a 121-layer network, etc.

Future work may include applying other kinds of image processing filters. New filters to assist such findings may also be a topic of interest. The utilized five filters were used with fixed parameters. Further work is required in fine tuning the parameters of these filters to gain further enhancements to accuracy.